\documentclass[twoside]{ilcws08}
\usepackage[latin1]{inputenc}
\usepackage[dvips]{graphicx,epsfig,color}
\usepackage{wrapfig,rotating}
\usepackage{amssymb,amsmath,array}

\newcommand{\zh}{$Z_{\mathrm{H}}$}
\newcommand{\ah}{$A_{\mathrm{H}}$}
\newcommand{\wh}{$W_{\mathrm{H}}$}
\newcommand{\whpm}{$W_{\mathrm{H}}^{\pm}$}
\newcommand{\wpm}{$W^{\pm}$}

\newcommand{\zhah}{$A_{\mathrm{H}}Z_{\mathrm{H}}$}
\newcommand{\whwh}{$W_{\mathrm{H}}^{+}W_{\mathrm{H}}^{-}$}

\newcommand{\zhahahahbb}{$A_{\mathrm{H}}Z_{\mathrm{H}} \to A_{\mathrm{H}} A_{\mathrm{H}} bb$}
\newcommand{\nnhnnbb}{$\nu \nu h \to \nu \nu bb$}
\newcommand{\zhnnbb}{$Zh \to \nu \nu bb$}
\newcommand{\zznnbb}{$ZZ \to \nu \nu bb$}
\newcommand{\nnznnbb}{$\nu \nu Z \to \nu \nu bb$}
\newcommand{\gzgbb}{$\gamma Z \to \gamma bb$}

\newcommand{\misspt}{$^{\mathrm{miss}}p_{\mathrm{T}}$}
\newcommand{\pt}{$p_{\mathrm{T}}$}

\newcommand{\wwqqqq}{$W^{+}W^{-} \to qqqq$}
\newcommand{\nnwwnnqqqq}{$\nu\bar{\nu}W^{+}W^{-} \to \nu\bar{\nu} qqqq$}
\newcommand{\eewweeqqqq}{$e^{+}e^{-} W^{+}W^{-} \to e^{+}e^{-} qqqq$}

\newcommand{\enwzenqqqq}{$e \nu_{e} W Z \to e \nu_{e} qqqq$}
\newcommand{\eeahzh}{$e^{+} e^{-} \to A_{\mathrm{H}} Z_{\mathrm{H}}$}
\newcommand{\eewhwh}{$e^{+} e^{-} \to W_{\mathrm{H}}^{+} W_{\mathrm{H}}^{-}$}
\newcommand{\whwhahahqqqq}{$W_{\mathrm{H}}^+W_{\mathrm{H}}^- \rightarrow A_{\mathrm{H}}A_{\mathrm{H}}qqqq$}
\newcommand{\zhzhahahhh}{$Z_{\mathrm{H}}Z_{\mathrm{H}} \rightarrow A_{\mathrm{H}}A_{\mathrm{H}} h h$}

\begin{document}
\title{Measurement of Heavy Gauge Bosons in Little Higgs Model with T-parity at ILC} 
\author{
Yosuke Takubo$^1$, 
Eri Asakawa$^2$,
Masaki Asano$^1$,
Keisuke Fujii$^3$,
Tomonori Kusano$^1$, \\
Shigeki Matsumoto$^4$,
Rei Sasaki$^1$, and
Hitoshi Yamamoto$^1$
\vspace{.3cm}\\
1- Department of Physics, Tohoku University, Sendai, Japan \\
2- Institute of Physics, Meiji Gakuin University, Yokohama, Japan \\
3- High Energy Accelerator Research Organization (KEK), Tsukuba, Japan \\
4- Department of Physics, University of Toyama, Toyama, Japan 
}

\maketitle

\begin{abstract}
The Littlest Higgs Model with T-parity is one of the attractive candidates 
of physics beyond the Standard Model. 
One of the important predictions of the model is the existence of new heavy gauge bosons, 
where they acquire mass terms through the breaking of global symmetry 
necessarily imposed on the model. 
The determination of the masses are, hence, quite important to test the model. 
In this paper, the measurement accuracy of the heavy gauge bosons at ILC is reported.
\end{abstract}

\section{Introduction}
There are a number of scenarios for new physics beyond the Standard Model. 
The most famous one is the supersymmetric scenario. 
Recently, alternative one called the Little Higgs scenario has been proposed
\cite{Arkani-Hamed:2001nc, Arkani-Hamed:2002qy}. 
In this scenario, the Higgs boson is regarded as a pseudo Nambu-Goldstone boson 
associated with a global symmetry at some higher scale. 
A $Z_2$ symmetry called T-parity is imposed on the models 
to satisfy constraints from electroweak precision measurements
\cite{Cheng:2003ju, Cheng:2004yc, Low:2004xc}. 
Under the parity, new particles are assigned to be T-odd (i.e. with a T-parity of $-1$), 
while the SM particles are T-even. 
The lightest T-odd particle is stable and provides a good candidate for dark matter. 
In this article, we focus on the Littlest Higgs model with T-parity 
as a simple and typical example of models implementing 
both the Little Higgs mechanism and T-parity. 

In order to test the Little Higgs model, 
precise determinations of properties of Little Higgs partners are mandatory, 
because these particles are directly related to the cancellation of 
quadratically divergent corrections to the Higgs mass term. 
In particular, measurements of heavy gauge boson masses,
Little Higgs partners for gauge bosons, are quite important. 
Since heavy gauge bosons acquire mass terms through the breaking of the global symmetry, 
precise measurements of their masses allow us to determine 
the most important parameter of the model, 
namely the vacuum expectation value of the breaking. 

We studied the measurement accuracy of masses of the heavy gauge bosons
at the international linear collider (ILC).
In addition, the sensitivity to the vacuum expectation value (f) was estimated.
In this paper, the status of the study is shown, 
and the detail of this study is described in \cite{prd}.

\section{Representative point and target mode}
In order to perform a numerical simulation at ILC, 
we need to choose a representative point 
in the parameter space of the Littlest Higgs model with T-parity. 
Firstly, the model parameters should satisfy the current electroweak precision data. 
In addition, the cosmological observation of dark matter relics also gives important information. 
Thus, we consider not only the electroweak precision measurements 
but also the WMAP observation \cite{Komatsu:2008hk} to choose a point in the parameter space.
We have selected a representative point 
where Higgs mass and $f$ are 134 GeV and 580 GeV, respectively.
At the representative point, we have obtained $\Omega_{\rm DM} h^2$ of 1.05. 
The masses of the heavy gauge bosons are 
($M_{A_{\mathrm{H}}}$, $M_{W_{\mathrm{H}}}$, $M_{Z_{\mathrm{H}}}$) = (81.9 GeV, 368 GeV, 369 GeV),
where \ah, \zh, and \wh~are the Little Higgs partners of a photon, Z boson, and W boson, respectively.
Here, \ah~plays the role of dark matter in this model \cite{Hubisz:2004ft, Asano:2006nr}.
Since all the heavy gauge bosons are lighter than 500 GeV, 
it is possible to generate them at ILC.

\begin{table}[t]
  \center{
    \begin{tabular}{|c||c|c|c|c|}
      \hline
      $\sqrt{s}$ &
      $e^+e^- \rightarrow A_{\mathrm{H}}Z_{\mathrm{H}}$ &
      $e^+e^- \rightarrow Z_{\mathrm{H}}Z_{\mathrm{H}}$ &
      $e^+e^- \rightarrow W_{\mathrm{H}}^+W_{\mathrm{H}}^-$ \\
      \hline
      500 GeV & 1.91 (fb) & --- & --- \\
      \hline
      1 TeV & 7.42 (fb) & 110 (fb) & 277 (fb) \\
      \hline
    \end{tabular}
  }
  \caption{\small Cross sections for the production of heavy gauge bosons.}
  \label{table:Xsections}
\end{table}

\begin{wrapfigure}{r}{0.4\columnwidth}
\centerline{\includegraphics[width=0.37\columnwidth]{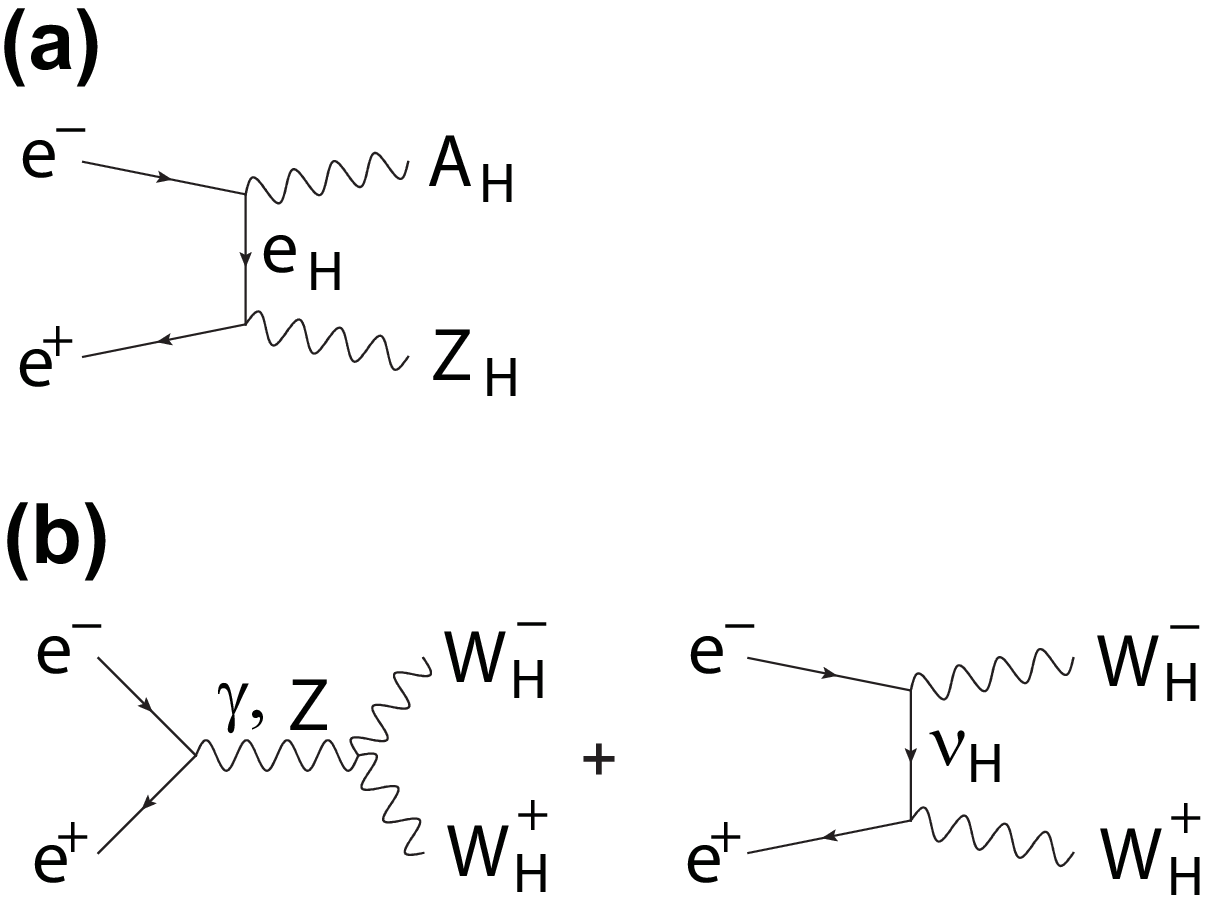}}
\caption{Diagrams for signal processes; (a) \eeahzh~and (b) \eewhwh.}
\label{fig:diagram}
\end{wrapfigure}

There are four processes whose final states consist of two heavy gauge bosons: 
$e^+e^- \rightarrow$ $A_{\mathrm{H}}A_{\mathrm{H}}$, $A_{\mathrm{H}}Z_{\mathrm{H}}$, 
$Z_{\mathrm{H}}Z_{\mathrm{H}}$, and $W_{\mathrm{H}}^+ W_{\mathrm{H}}^-$. 
The first process is undetectable, thus not considered in this article. 
The cross sections of the other processes are shown in Table \ref{table:Xsections}. 
Since $m_{A_{\mathrm{H}}} + m_{Z_{\mathrm{H}}}$ is less than 500 GeV, 
\zhah~can be produced at the $\sqrt{s} = 500$ GeV. 
At $\sqrt{s} = 1$ TeV, we can observe \whwh~with large cross section.
We, hence, concentrate on \eeahzh~at $\sqrt{s} = 500$ GeV and \eewhwh~at $\sqrt{s} = 1$ TeV. 
Feynman diagrams for the signal processes are shown in Fig. \ref{fig:diagram}. 
Note that \zh~decays into $A_{\mathrm{H}} h$, and \whpm~decays into $A_{\mathrm{H}}W^\pm$ 
with almost 100\% branching fractions.

\section{Simulation tools}
We have used MadGraph \cite{madgraph} to generate \eeahzh~at $\sqrt{s} =$ 500 GeV, 
while \eewhwh~at $\sqrt{s} =$ 1 TeV and all the standard model events 
have been generated by Physsim \cite{physsim}. 
We ignored the initial- and final-state radiation, 
beamstrahlung, and the beam energy spread for study of \eeahzh~at $\sqrt{s} =$ 500 GeV,
whereas their effects were considered for study of \eewhwh~at $\sqrt{s} =$ 1 TeV
where the beam energy spread is set to 0.14\% for the electron beam 
and 0.07\% for the positron beam.
The finite crossing angle between the electron and positron beams was assumed to to be zero. 
In both event generators, 
the helicity amplitudes were calculated using the HELAS library \cite{helas}, 
which allows us to deal with the effect of gauge boson polarizations properly. 
Parton showering and hadronization have been carried out by using PYTHIA6.4 \cite{pythia}, 
where final-state tau leptons are decayed by TAUOLA \cite{tauola} 
in order to handle their polarizations correctly.
The generated Monte Carlo events have been passed to a detector simulator 
called JSFQuickSimulator, 
which implements the GLD geometry and other detector-performance related parameters 
\cite{glddod}. 

\section{Analysis}
In this section, we present simulation and analysis results  
for heavy gauge boson productions. The simulation has been performed 
at $\sqrt{s} =$ 500 GeV for the \zhah~production and at $\sqrt{s} =$ 1 TeV 
for the \whwh~production 
with an integrated luminosity of 500 fb$^{-1}$.

\subsection{\eeahzh~at 500 GeV}
\begin{wrapfigure}{r}{0.35\columnwidth}
\centerline{\includegraphics[width=0.33\columnwidth]{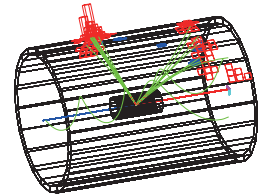}}
\caption{A typical event of \zhah~in the simulator.}
\label{fig:evtdsp}
\end{wrapfigure}

\ah~and \zh~are produced with the cross section of 1.9 fb 
at the center of mass energy of 500 GeV. 
Since \zh~decays into \ah~and the Higgs boson, 
the signature is a single Higgs boson in the final state, 
mainly 2 jets from $h \to b\bar{b}$ (with a 55\% branching ratio). 
We, therefore, define \zhahahahbb~as our signal event. 
For background events, contribution from light quarks was not taken into account 
because such events can be rejected to negligible level 
after requiring the existence of two $b$-jets, 
assuming a $b$-tagging efficiency of 80\% for $b$-jets with 15\% probability 
to misidentify a $c$-jet as a $b$-jet. 
This $b$-tagging performance was estimated by the full simulation, 
assuming a typical ILC detector. 
Signal and background processes considered in this analysis are summarized 
in Table \ref{tb:zhevlst}. 
Figure \ref{fig:evtdsp} shows a typical \zhah~event seen in the detector simulator.

\begin{table}
\center{
\begin{tabular}{l|r|r|r}
\hline
Process  & Cross sec. [fb] & \# of events & \# of events after all cuts \\
\hline
\zhahahahbb & 1.05  & 525     & 272   \\
\nnhnnbb    & 34.0  & 17,000  & 3,359 \\
\zhnnbb     & 5.57  & 2,785   & 1,406 \\
$tt \to WWbb$ & 496 & 248,000 & 264   \\
\zznnbb     & 25.5  & 12,750  & 178   \\
\nnznnbb    & 44.3  & 22,150  & 167   \\
\gzgbb      & 1,200 & 600,000 & 45    \\
\hline
\end{tabular}
}
\caption{\small Signal and backgrounds processes considered in the \zhah~analysis.}
\label{tb:zhevlst}
\end{table}

The clusters in the calorimeters are combined to form a jet 
if the two clusters satisfy $y_{ij} < y_{\mathrm{cut}}$. $y_{ij}$ is defined as
\begin{equation}
 y_{ij} = \frac{2 E_{i} E_{j} (1 - \cos \theta_{ij})}{E_{\mathrm{vis}}^{2}},
\end{equation}
where $\theta_{ij}$ is the angle between momenta of two clusters, 
$E_{i(j)}$ are their energies, and $E_{\mathrm{vis}}$ is the total visible energy. 
All events are forced to have two jets by adjusting $y_{\mathrm{cut}}$. 
We have selected events with the reconstructed Higgs mass in a window of $100-140$ GeV. 
Since Higgs bosons coming from the $WW$ fusion process have 
the transverse momentum (\pt) mostly below W mass,
\pt~is required to be above 80 GeV in order to suppress the \nnhnnbb~background.
Finally, multiplying the efficiency of double $b$-tagging ($0.8 \times 0.8 = 0.64$), 
we are left with 272 signal and 5,419 background events as shown in Table \ref{tb:zhevlst}, 
which corresponds to a signal significance of 3.7 ($= 272/\sqrt{5419}$) standard deviations. 
The indication of the new physics signal can hence be obtained at $\sqrt{s} = 500$ GeV.

\begin{figure}
\begin{center}
\includegraphics[width=10cm]{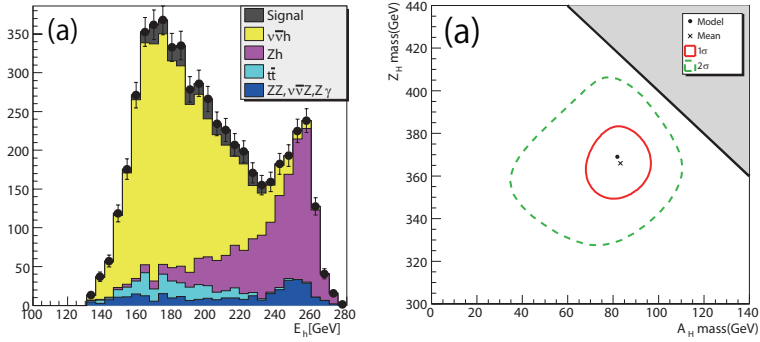}
\end{center}
\caption{(a)Energy distribution of the reconstructed Higgs bosons 
with remaining backgrounds after the mass cut.
(b) Probability contours corresponding to 1- and 2-$\sigma$ deviations from the best fit point
in the \ah~and \zh~mass-plane. The shaded area shows the unphysical region of $m_{A_{\mathrm{H}}} + m_{Z_{\mathrm{H}}} > 500$ GeV.}
\label{fig:hene}
\end{figure}

The masses of \ah~and \zh~bosons can be estimated from the edges of the distribution 
of the reconstructed Higgs boson energies. 
This is because the maximum and minimum Higgs boson energies 
($E_{\mathrm{max}}$ and $E_{\mathrm{min}}$) are written in terms of these masses,

\begin{eqnarray}
 E_{\mathrm{max}}
 &=& 
 \gamma_{Z_{\mathrm{H}}} E^{\ast}_{h}
 + 
 \beta_{Z_{\mathrm{H}}} \gamma_{Z_{\mathrm{H}}} p^{\ast}_{h},
 \nonumber \\ 
 E_{\mathrm{min}}
 &=& 
 \gamma_{Z_{\mathrm{H}}} E^{\ast}_{h}
 - 
 \beta_{Z_{\mathrm{H}}} \gamma_{Z_{\mathrm{H}}} p^{\ast}_{h},  
 \label{eq:eedge}
\end{eqnarray}
where $\beta_{Z_{\mathrm{H}}} (\gamma_{Z_{\mathrm{H}}})$ is the $\beta (\gamma)$ factor 
of the \zh~boson in the laboratory frame, while $E^{\ast}_{h} 
(p_{h}^{\ast})$ is the energy (momentum) of the Higgs boson 
in the rest frame of the \zh~boson. Note that $E^{\ast}_{h}$ is given 
as $(M_{Z_{\mathrm{H}}}^2 + M_h^2 - M_{A_{\mathrm{H}}}^2)/(2M_{Z_{\mathrm{H}}})$.

Figure \ref{fig:hene}(a) shows the energy distribution of the reconstructed Higgs bosons 
with remaining backgrounds. 
The background events are subtracted from Fig. \ref{fig:hene}(a),
assuming that the background distribution can be understand completely.
Then, the endpoints, $E_{\mathrm{max}}$ and $E_{\mathrm{min}}$, 
have been estimated by fitting the distribution with a line shape determined 
by a high statistics signal sample. 
The fit resulted in $m_{A_{\mathrm{H}}}$ and $m_{Z_{\mathrm{H}}}$ 
to be $83.2 \pm 13.3$ GeV and $366.0 \pm 16.0$ GeV, respectively, 
which should be compared to their true values: 81.85 GeV and 368.2 GeV. 
Figure \ref{fig:hene}(b) shows the probability contours for the masses of \ah~and \zh.

Since the masses of the heavy gauge bosons are from the vacuum expectation value ($f$), 
$f$ can be determined by fitting the energy distribution of the reconstructed Higgs bosons. 
Then, $f$ was determined to be $f = 576.0 \pm 25.0$ GeV.

\subsection{\eewhwh~at 1 TeV}
$W_{\mathrm{H}}^+ W_{\mathrm{H}}^-$ production has large cross section (277 fb) 
at ILC with $\sqrt{s} = 1$ TeV. 
Since \whpm~decays into \ah~and \wpm~ with the 100\% branching ratio, 
analysis procedure depends on the $W$ decay modes. 
In this analysis, we have used 4-jet final states from hadronic decays of two $W$ bosons, 
$W_{\mathrm{H}}^+W_{\mathrm{H}}^- \rightarrow A_{\mathrm{H}} A_{\mathrm{H}} qqqq$. 
Signal and background processes considered in the analysis are summarized 
in Table \ref{tb:whevlst}.

\begin{table}
\center{
\begin{tabular}{l|r|r|r}
\hline
Process       & cross sec. [fb] & \# of events & \# of events after all cuts \\ \hline
\whwhahahqqqq & 106.5           & 53,258       & 37,560          \\
\wwqqqq       & 1773.5          & 886,770      & 306             \\
\eewweeqqqq   & 464.9           & 232,442      & 23              \\
\enwzenqqqq   & 25.5            & 12,770       & 3,696           \\
\zhzhahahhh   & 99.5            & 49,757       & 3,351           \\
\nnwwnnqqqq   & 6.5             & 3,227        & 1,486           \\
\hline
\end{tabular}
}
\caption{\small Signal and background processes considered 
in the $W_{\mathrm{H}}^+ W_{\mathrm{H}}^-$ analysis.}
\label{tb:whevlst}
\end{table}

All events have been reconstructed as 4-jet events by adjusting the cut on y-values. 
In order to identify the two $W$ bosons from \whpm~decays, 
two jet-pairs have been selected so as to minimize a $\chi^2$ function,
\begin{equation}
\chi^2
= 
(^{\mathrm{rec}}\mathrm{M}_{W1} -~^{\mathrm{tr}}\mathrm{M}_{W})^{2}/\sigma_{\mathrm{M}_{W}}^{2} 
+ 
(^{\mathrm{rec}}\mathrm{M}_{W2} -~^{\mathrm{tr}}\mathrm{M}_{W})^{2}/\sigma_{\mathrm{M}_{W}}^{2},
\end{equation}
where $^{\mathrm{rec}}\mathrm{M}_{W1(2)}$ is the invariant mass 
of the first (second) 2-jet system paired 
as a $W$ candidate, $^{\mathrm{tr}}\mathrm{M}_{W}$ is the true $W$ mass (80.4 GeV), 
and $\sigma_{\mathrm{M}_{W}}$ is the resolution for the $W$ mass (4 GeV). 
We required $\chi^2 < 26$ to obtain well-reconstructed events. 
Since \ah~bosons escape from detection resulting in a missing momentum, 
the missing transverse momentum (\misspt) of the signal peaks at around 175 GeV. 
We have thus selected events with \misspt~above 84 GeV.
Then, the reconstructed W energy is required to be between 0 GeV to 500 GeV. 
The numbers of events after the selection cuts are shown in Table \ref{tb:whevlst}. 
The number of remaining background events is much smaller than that of the signal.

\begin{figure}
\begin{center}
\includegraphics[width=10cm]{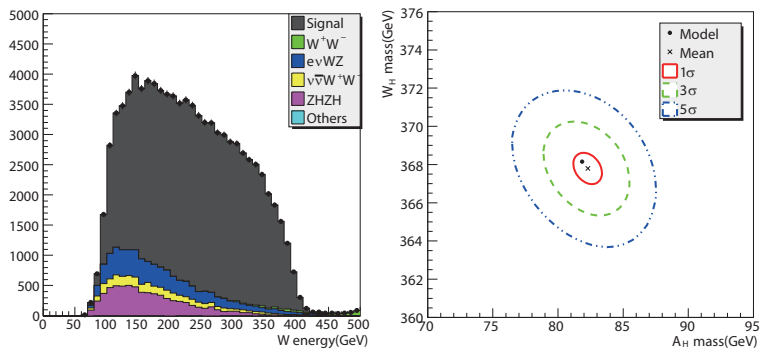}
\end{center}
\caption{(a) The energy distribution of the reconstructed $W$ bosons 
with remaining backgrounds after the selection cuts.
(b) Probability contours corresponding to 1-, 3-, and 5-$\sigma$ deviations
in the \ah~and \wh~mass-plane.}
\label{fig:wene}
\end{figure}

As in the case of the \zhah~production, 
the masses of \ah~ and \wh~bosons can be determined from the edges of the $W$ energy distribution. 
Figure \ref{fig:wene}(a) shows the energy distribution of the reconstructed $W$ bosons. 
After subtracting the backgrounds from Fig.\ref{fig:wene}(a), 
the distribution has been fitted with a line shape function. 
The fitted masses of \ah~and \wh~ bosons are $82.29 \pm 1.10$ GeV and $367.8 \pm 0.8$ GeV, 
respectively, which are to be compared to their input values: 81.85 GeV and 368.2 GeV. 
Figure \ref{fig:wene}(b) shows the probability contours for the masses of \ah~and \wh~at $1$ TeV. 
The mass resolution improves dramatically at $\sqrt{s} = 1 $ TeV, 
compared to that at $\sqrt{s} = 500$ GeV.
Then, $f = 579.7 \pm 1.1$ GeV was obtained by fitting the energy distribution of
the reconstructed W bosons. 

\section{Summary} \label{sec:summary}
The Littlest Higgs Model with T-parity is one of the attractive candidates 
of physics beyond the Standard Model since it solves both the little hierarchy 
and dark matter problems simultaneously. 
One of the important predictions of the model is the existence of new heavy gauge bosons, 
where they acquire mass terms through the breaking of global symmetry 
necessarily imposed on the model. 
The determination of the masses are, hence, quite important to test the model. 

We have performed Monte Carlo simulations 
in order to estimate measurement accuracy of the masses of the heavy gauge bosons at ILC.
At ILC with $\sqrt{s} = 500$ GeV, it is possible to produce \ah~ and \zh~bosons. 
Here, we can observe the excess by \zhah~events 
in the Higgs energy distribution with the statistical significance of 3.7-sigma. 
Furthermore, the masses of these bosons can be determined with accuracies of 16.2\% 
for \ah~and 4.3\% for \zh.
Once ILC energy reaches $\sqrt{s}=$ 1 TeV, 
the process $e^+e^- \rightarrow W_{\mathrm{H}}^+W_{\mathrm{H}}^-$ opens. 
Since the cross section of the process is large, 
the masses of \wh~and \ah~can be determined as accurately as 1.3\% and 0.2\%, respectively.
Then, the vacuum expectation value, $f$, can be determined 
with accuracy of 4.3\% at $\sqrt{s}=$ 500 GeV and 0.2\% at 1 TeV. 


\section{Acknowledgments}
The authors would like to thank all the members of the ILC physics subgroup
\cite{softg} for useful discussions. 
This study is supported in part by the Creative Scientific Research Grant
No. 18GS0202 of the Japan Society for Promotion of Science, and
Dean's Grant for Exploratory Research in Graduate School of Science of Tohoku University.


\begin{footnotesize}




\begin{thebibliography}{99}

\bibitem{Arkani-Hamed:2001nc}
  N.~Arkani-Hamed, A.~G.~Cohen and H.~Georgi,
  Phys.\ Lett.\ B {\bf 513} (2001) 232;

\bibitem{Arkani-Hamed:2002qy}
  N.~Arkani-Hamed, A.~G.~Cohen, E.~Katz and A.~E.~Nelson,
  JHEP {\bf 0207} (2002) 034.

\bibitem{Cheng:2003ju}
  H.~C.~Cheng and I.~Low,
  JHEP {\bf 0309} (2003) 051.

\bibitem{Cheng:2004yc}
  H.~C.~Cheng and I.~Low,
  JHEP {\bf 0408} (2004) 061.

\bibitem{Low:2004xc}
  I.~Low,
  JHEP {\bf 0410} (2004) 067.

\bibitem{prd} E. Asakawa, Phys. Rev. D79, 075013, (2009).

\bibitem{Komatsu:2008hk}
  E.~Komatsu {\it et al.}  [WMAP Collaboration],
  arXiv:0803.0547 [astro-ph].

\bibitem{Hubisz:2004ft}
  J.~Hubisz and P.~Meade,
  Phys.\ Rev.\ D {\bf 71} (2005) 035016,
  (For the correct paramter region consistent with the WMAP observation,
   see the figure in the revised vergion, hep-ph/0411264v3).

\bibitem{Asano:2006nr}
  M.~Asano, S.~Matsumoto, N.~Okada and Y.~Okada,
  Phys.\ Rev.\  D {\bf 75} (2007) 063506;

\bibitem{madgraph} http://madgraph.hep.uiuc.edu/.
\bibitem{physsim} http://acfahep.kek.jp/subg/sim/softs.html.
\bibitem{helas} H. Murayama, I. Watanabe, K. Hagiwara, KEK-91-11, (1992) 184.
\bibitem{pythia} T. Sj$\dot{\mathrm{o}}$strand, \emph{Comp, Phys. Comm.} {\bf 82} (1994) 74.
\bibitem{tauola} http://wasm.home.cern.ch/wasm/goodies.html.
\bibitem{glddod} GLD Detector Outline Document, arXiv:physics/0607154.
\bibitem{softg} http://www-jlc.kek.jp/subg/physics/ilcphys/.
\end{thebibliography}
%

\end{footnotesize}


\end{document}